\documentclass[fleqn,10pt]{wlscirep}

\usepackage[utf8]{inputenc}
\usepackage[T1]{fontenc}

\usepackage{fdsymbol}

\usepackage[export]{adjustbox}
\usepackage{textcomp}
\usepackage[separate-uncertainty,multi-part-units=single]{siunitx}

\DeclareSIUnit\degree{deg}
\DeclareSIUnit\trial{trial}
\DeclareSIUnit\year{yr}
\newcommand{\vect}[1]{\mathbf{#1}}

\title{Haptic human-human interaction does not improve individual visuomotor adaptation}
\author[1,2,*]{Niek Beckers}
\author[1]{Edwin van Asseldonk}
\author[1,3]{Herman van der Kooij}
\affil[1]{Department of Biomechanical Engineering, University of Twente, Enschede, The Netherlands}
\affil[2]{Cognitive Robotics, Delft University of Technology, Delft, The Netherlands}
\affil[3]{Department of Biomechanical Engineering, Delft University of Technology, Delft, The Netherlands}
\affil[*]{Correspondence should be addressed to N.B. (niekbeckers@gmail.com)}

\begin{abstract}
Haptic interaction between two humans, for example, a physiotherapist assisting a patient regaining the ability to grasp a cup, likely facilitates motor skill acquisition. Haptic human-human interaction has been shown to enhance individual performance improvement in a tracking task with a visuomotor rotation perturbation. These results are remarkable given that haptically assisting or guiding an individual rarely benefits their individual improvement when the assistance is removed. We, therefore, replicated a study that reported that haptic interaction between humans was beneficial for individual improvement for tracking a target in a visuomotor rotation perturbation. In addition, we tested the effect of more interaction time and a stronger haptic coupling between the partners on individual improvement in the same task. We found no benefits of haptic interaction on individual improvement compared to individuals who practised the task alone, independent of interaction time or interaction strength.
\end{abstract}

\begin{document}

\flushbottom
\maketitle
\thispagestyle{empty}

%%
%% Introduction
%%
\section*{Introduction}

While working to improve their motor abilities, such as walking or grasping a cup, patients often rely on physical assistance of their physiotherapist. 
Joint action, like physical assistance, is a crucial way through which we learn new skills or transfer knowledge to others \cite{Sebanz:2006bd,tomasello_understanding_2005}. 
Here, we study the effect of haptic interaction, in which two partners exchange forces while performing a joint task, on motor skill acquisition.

To date, the number of studies on the effect of haptic interaction between humans on individual motor skill acquisition is limited and their results are dissimilar. 
Ganesh et al.\cite{Ganesh:2014sr} were, to our knowledge, one of the first to report that haptic interaction between two partners resulted in better \emph{individual} motor performance after haptic interaction compared to participants who never interacted. 
They intermittently coupled two partners with a compliant spring generated by a dual-robot interface while they tracked the same continuously moving target in a challenging visuomotor rotation perturbation. 
A visuomotor rotation perturbation is a motor adaptation paradigm in which the visual feedback of the arm movement is rotated with respect to the actual arm movement. Motor performance is initially decreased when the visuomotor rotation is introduced, but people consistently improve performance with practice in a visuomotor rotation by compensating for the visuomotor rotation, a process referred to as motor adaptation \cite{Shadmehr:2010fi}. 
Ganesh et al.\cite{Ganesh:2014sr} analysed to what extent intermittent haptic interaction influenced each participant's individual tracking performance in trials in which the participants performed the task alone. The individual performance was compared to a group who never interacted.
They showed that intermittent haptic interaction improved the participant's \emph{individual} motor improvement significantly \emph{more} and, although not explicitly mentioned by the authors, initially \emph{faster} compared to someone who practised the task alone. These results are encouraging and have potential high impact for designing robot-assisted motor skill acquisition algorithms, for example for physical rehabilitation applications.

However, more recent studies report no or context- or task-specific benefits of haptic interaction on individual motor skill acquisition.
Takagi et al. \cite{Takagi:2017kv,Takagi:2018ku} found no benefit on improvement rates in tracking tasks without visual perturbation, though their tracking task might have been not challenging enough to elicit significant differences in individual motor improvement.
In an earlier study, we found that haptic interaction did not improve individual motor adaptation to a velocity-dependent force field while tracking a continuously moving target \cite{Beckers:2018bi}.  
Van der Wel et al.\cite{vanderWel:2011hg} reported that haptically interacting partners learned a novel coordination task (balancing a stick) just as quickly as individuals performing the task alone.
Using a tracking task and the same visuomotor perturbation as Ganesh et al.\cite{Ganesh:2014sr}, Kager et al.\cite{kager2019effect} found no significant effect of haptic interaction on final individual motor performance. However, their study has a few important differences to Ganesh et al.\cite{Ganesh:2014sr}: their tracking task was less challenging, they used a different haptic interaction paradigm and they only tested a limited number of participants.
Lastly, practising a challenging reaching task with nonlinear dynamics while being haptically connected to an expert appeared to be less beneficial for subsequent individual motor performance than being connected to a partner with similar initial skill level \cite{AvilaMireles:2017cf}. Unfortunately, the final individual skill levels of their interaction participants were not compared to a solo group that practised the task without interaction for the same amount of time. 
Based on these results -- although obtained using different motor tasks -- we question \emph{whether} haptic human-human interaction indeed benefits an individual's visuomotor adaptation as found by Ganesh et al.\cite{Ganesh:2014sr}.

It is also unknown \emph{why} haptic interaction would improve individual motor adaptation. Adapting to a visuomotor rotation is predominantly driven by errors between planned and actual movements \cite{Shadmehr:2010fi}. The larger the experienced error, the more participants compensate for those errors in subsequent movements \cite{Wei:2009bi}. Reducing a participant's movement errors while they learn a new motor skill using robot-generated haptic assistance rarely transferred into improved subsequent individual performance compared to participants who received no haptic assistance \cite{Winstein:1994gf,vanAsseldonk:2009do,OMalley:2006dv,MarchalCrespo:2014ec,Heuer:2015tb}.
Similarly, several studies on haptic human-human interaction consistently showed that tracking errors are significantly smaller \emph{during} haptic interaction compared to performing the tracking task alone, particularly if you interact with a partner who is better than you \cite{Ganesh:2014sr,Takagi:2017kv,Takagi:2018ku,AvilaMireles:2017cf,kager2019effect,Beckers:2018bi}. 
Combining these observations, we would expect little to no benefit of haptic human-human interaction on individual visuomotor adaptation.

Still, haptic interaction could be a valuable means through which partners transfer skill or learn from each other. Visually observing another person learn a motor skill promotes the observer's own learning \cite{Sebanz:2006bd,Mattar:2004bv}. Similarly, haptic interaction could enable an individual to imitate or observe the actions of their partner through haptics, facilitating their own motor skill acquisition.  
Also, haptically interacting pairs can communicate intentions or coordinate specialised roles through the interaction force\cite{Reed:2008ie,Groten:2013jk,Sawers:2017fj,Mojtahedi:2017jl,Kucukyilmaz:2013ft}; partners could adopt teaching roles, coaching each other on how to account for the visuomotor rotation, potentially speeding up their visuomotor adaptation. 
However, coordinating roles or communicating intentions would likely take additional time on top of the motor adaptation task. A joint action study showed that groups can learn to coordinate actions, but the group coordination process only occurred gradually, which could hinder task performance compared to individuals \cite{knoblich_action_2003}.
Still, although the aforementioned mechanisms could facilitate motor adaptation, we believe it is unlikely that participants who were na\"ive to the haptic interaction -- like in Ganesh et al.\cite{Ganesh:2014sr} --  were able to use any of these mechanisms to improve their short-term motor adaptation, especially in the early stages of the adaptation process. 

Our goal is to investigate whether haptic interaction with a partner who is practising the same tracking task indeed results in more and faster individual performance improvement in a visuomotor rotation perturbation. In addition to replicating the experiment by Ganesh et al.\cite{Ganesh:2014sr}, we also investigate whether the amount of interaction time (i.e. more trials in which the partners interact) and the strength of the haptic coupling affect individual improvement. 
The participants in Ganesh et al.\cite{Ganesh:2014sr} interacted intermittently in half of the trials. If haptic interaction indeed benefits individual motor improvement, more interaction time could, on the one hand, increase these benefits. On the other hand, as haptic interaction reduces tracking error -- a key training signal for visuomotor adaptation -- practising the task always connected to a partner could yield no individual improvement benefits.
To further test the hypothesised effect of the reduced experienced tracking error during interaction on visuomotor adaptation, we also tested a group who interacted through a stronger haptic coupling. A stronger coupling has been show to reduce tracking errors significantly more than a weaker coupling, in particular for the inferior-performing partner \cite{Takagi:2018ku}, which could affect individual motor improvement\cite{vanAsseldonk:2009do}. 
Lastly, as the partner's skill level can significantly affect a participant's \emph{individual} performance \cite{AvilaMireles:2017cf}, we also analysed the effect of the partner's relative initial skill level on the participant's improvement and improvement rates.

Contrary to Ganesh et al. \cite{Ganesh:2014sr}, we found no benefit of haptic human-human interaction on individual visuomotor adaptation in a tracking task compared to individuals who practised the task alone. Haptic interaction did not result in more individual improvement or faster improvement. Increasing the amount of interaction time or interaction strength also did not improve or impede individual improvement.

\section*{Methods}

Eighty healthy participants were recruited: 40 women and 40 men; age \SI{22.0\pm2.1}{\year}; all except three participants were right-handed according to the Edinburgh handedness inventory \cite{Oldfield:1971kz}. The participants were equally distributed over the four experiment groups that we will describe later on. The participants had no prior experience with studies involving haptic human-human interaction or visuomotor rotation adaptation paradigms. 
An assessment of the study by the Medical Ethical Review Board of the University of Twente (METC Twente) showed that the study posed minimal risk to the participants and therefore under Dutch law did not need full ethical review. All participants provided written informed consent. All participants received a compensation for their participation, independent of their performance or whether they completed the study. The experiment lasted approximately two hours.

\subsection*{Dual-robot interface}
\begin{figure}[ht]
	\centering
	% \tikzsetnextfilename{Figure-1-Beckers}
	% \input{./figures/tikz-src/bros-setup-schematic}
    \includegraphics{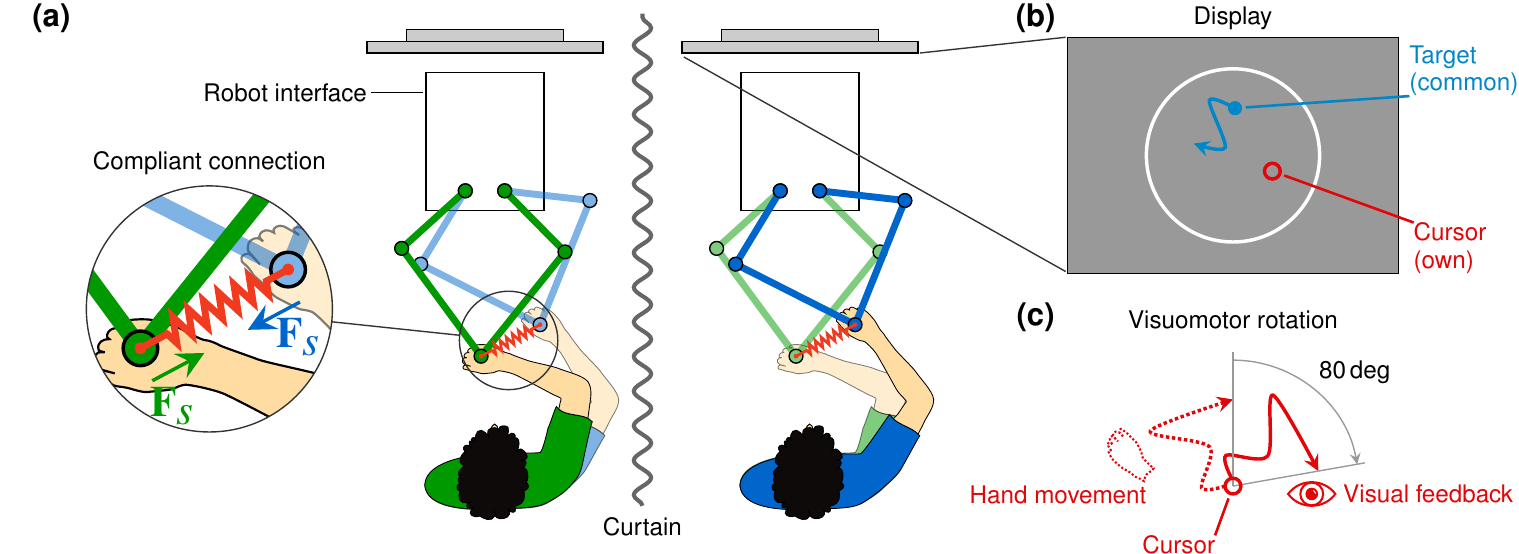}
	\caption{Dual-robot interface, display, and visuomotor rotation paradigm. \textbf{(a)} Each participant held the handle of one of the two identical robot interfaces. Visual feedback was presented on a display mounted in front of each participant. The robot interfaces could generate a compliant connection between the partners to enable haptic interaction. \textbf{(b)} Participants received visual feedback of their own cursor and a common target; they did not see their partner's cursor. \textbf{(c)} Visuomotor rotation: the visual feedback of the cursor was rotated clockwise with \SI{80}{\degree} with respect to the actual hand movement.
	}
	\label{fig:learning-bros-setup}
\end{figure}

The experiment was performed using a dual-robot interface (see Fig.~\ref{fig:learning-bros-setup}a). Participants held a handle at the endpoint of their own robot interface with their preferred hand. Each robot interface allowed hand movements in a planar circular workspace with a diameter of \SI{20}{\centi\meter}. 
Each participant had their own display that showed the workspace, the target and their own cursor that they could control by moving the robot interface's handle (see Fig.~\ref{fig:learning-bros-setup}b). Cursor movement was scaled to match the real-world movement of the handle. The coordinate frame of each robot interface was centred at the centre of the corresponding display. A panel obstructed the participant's view on their arm, hand and robot interface. A curtain separated the partners to prevent social interaction. 

\subsection*{Tracking task, visuomotor rotation paradigm, and haptic interaction paradigm}
The tracking task and haptic interaction paradigm were similar to the ones used by Ganesh et al.\cite{Ganesh:2014sr}. All participants tracked a continuously moving target with their own cursor as accurately as possible in trials with a duration of \SI{23}{\second} each. 
The target trajectory (in \si{\milli\meter}) was defined as a sum-of-sines (see Supplementary Methods for more details on the target signal design):
\begin{equation}
	\begin{aligned}
    x(t)= 28.7\sin&\left(0.94t-7.77\right) + 27.1\sin\left(1.26t-8.53\right)+ 23.5\sin\left(1.89t-4.36\right) + 18.0\sin\left(2.83t-3.79\right)\text{,}\\
    y(t)= 27.1\sin&\left(1.26t-0.71\right) + 25.3\sin\left(1.57t-3.45\right)+ 21.6\sin\left(2.20t+3.92\right) + 16.4\sin\left(3.14t+4.93\right)\textrm{.}
  \end{aligned}
  \label{eq:learning-trackingsignal}
\end{equation}
The tracking signal required hand movements over the robot interface's full circular workspace, an average velocity of \SI{7.9}{\centi\meter\per\second} and a maximum velocity of \SI{13.9}{\centi\meter\per\second}.
We generated the tracking signal using different time offsets $t_0$ for each trial, which was randomly chosen from a uniform distribution $\left(t \in\left[t_0,t_0+20\right] \si{\second}\textrm{, } t_0 \sim \mathcal{U}(0,20) \si{\second}\right)$ for each trial. As a result, the target started at different locations of the target trajectory in each trial to help keep the tracking task challenging and slightly different per trial.

The participants performed 84 trials divided over four blocks with \SI{15}{\second} of rest between trials and five minutes of rest between blocks. 
The first block served as a baseline block in which the participants tracked the moving target without the visuomotor rotation; hence the visual movements of the cursor matched the actual hand movements. 

We then introduced the same \SI{80}{\degree} visuomotor rotation as used by Ganesh et al.\cite{Ganesh:2014sr} in blocks 2, 3, and 4 to study the effect of haptic interaction on individual tracking improvement. The cursor movement on the display was rotated clockwise with \SI{80}{\degree} with respect to the participant's actual hand movement (see Fig.~\ref{fig:learning-bros-setup}c), which initially results in a mismatch between the expected cursor movement based on the hand movement and actual cursor movement. A visuomotor rotation initially degrades tracking performance, which then improves with practice, typically within a few hours, as the participants adapt to the different mapping between actual hand movements and the visual feedback of their hand movements \cite{Shadmehr:2010fi}.

The haptic interaction paradigm is the same as used by Ganesh et al.\cite{Ganesh:2014sr}. Two types of trials were used in the experiment: \emph{single} trials (S), in which the participants performed the tracking task alone, and \emph{connected} trials (C). 
In the connected trials the partners' hands were coupled through an compliant connection (see the detail in Fig.~\ref{fig:learning-bros-setup}a) with a force 
\begin{equation}
   \vect{F}_s= k_{s} \left(\vect{p}_p-\vect{p}_o\right) + b_{s} \left(\dot{\vect{p}}_p-\dot{\vect{p}}_o\right) % \textrm{.}
\end{equation}
generated by each robot interface. 
A participant would experience an interaction force $\vect{F}_s$ when he/she was at position $\vect{p}_o$ and their partner was at position $\vect{p}_p$.
The coordinate frames of the robot interfaces coincided, so that if both partners moved along the same trajectory, they would experience no interaction force, while if a partner moved away from the other partner, they both experienced a force pulling them toward each other.
The stiffness $k_s$ was set to \SI{120}{\newton\per\meter} (same as used by Ganesh et al.\cite{Ganesh:2014sr}) or \SI{250}{\newton\per\meter}, depending on the experiment group and the damping was set to $b_s=\SI{7}{\newton\per\meter}$.
The compliant connection stiffnesses were chosen such that the task required active tracking: participants could not completely relax and let the interaction force pull their hand passively along.

% instructions
We instructed all participants to track the target as accurately as possible using continuous and smooth movements; their goal was to minimise the tracking error as much as they could. 
We also explained the concept of a visuomotor rotation and made clear that their goal was to accurately track the target despite the rotation and not to estimate the magnitude of the rotation. 
We informed participants that the interaction forces they would sometimes experience ``involved external forces that would sometimes help the task and sometimes disturb it''. We did not provide explicit information about the haptic connection.
Participants were not allowed to verbally communicate during the experiment.

\subsection*{Experiment groups}

The eighty participants were equally divided over four groups: (1) a \emph{solo} group, (2) an `intermittent interaction' group (denoted by \emph{int. int.}), (3) a `stiff interaction' group (\emph{stiff int.}), and (4) a `continuous interaction' group (\emph{cont. int.}).  
All participants in each group performed the experiment in gender- and age-matched pairs.
The groups performed the same tracking task with the same visuomotor rotation for the same amount of trials. We only changed how often participants in a group interacted (i.e. the amount of connected trials versus the amount of single trials) or the strength of the coupling between the partners in the connected trials.

The \emph{solo} participants performed the experiment in pairs, but they were never connected. Solo participants thus performed performed the tracking task always alone; they only performed single trials (S). The solo participants served as a control group.

The \emph{intermittent interaction} group intermittently interacted through the connection with a stiffness of $k_s=\SI{120}{\newton\per\meter}$. Each block consisted of sequences of alternative single (S) and connected (C) trials, resulting in the same trial sequence for each of the four blocks: \{SCSCSCSCSCSCSCSCSCSCS\} per block. The first single trial of block 2 (i.e. the first single trial with the visuomotor rotation) was used to assess the individual initial tracking error in the visuomotor rotation, denoted by $E_{s,0}$. 
The intermittent interaction group is similar to the interaction group of Ganesh et al.\cite{Ganesh:2014sr} and was used for comparison to their results. We used the same connection stiffness ($k_s=\SI{120}{\newton\per\meter}$) and damping ($b_s=\SI{7}{\newton\second\per\meter}$). 
Our trial sequence differs from Ganesh et al.\cite{Ganesh:2014sr}, who used a semi-random sequence of single and connected trials. We chose to alternate the single and connected trials to have a consistent spacing of single trials (for comparison with the solo and other experiment groups) throughout the experiment to capture each participants motor improvement over time.

We increased the connection stiffness $k_s$ to \SI{250}{\newton\per\meter} for the \emph{stiff interaction} group to investigate the effect of higher interaction strength on individual motor improvement. Pilot tests showed that the higher connection stiffness resulted in stronger interaction forces compared to the intermittent interaction group, yet still allowed for independent movement. 
The stiff interaction group used the same alternating sequence of single and connected trials per block as the intermittent interaction group. Similarly, the first single trial in the visuomotor rotation blocks was used as the individual initial tracking error $E_{s,0}$.

The \emph{continuous interaction} group had different trial sequences per block compared to the int. int. and stiff int. groups. Block 1 consisted of single trials only. 
The partners were always connected in the visuomotor rotation blocks (blocks 2, 3 and the first half of block 4) to investigate the effect of more interaction time on individual motor improvement in the visuomotor rotation. Specifically, we used the the following trial sequence for this group:
[\{$21\times$S\},\{$21\times$C\},\{$21\times$C\},\{CCCCCCCCC SSSSSSSSSSSS\}].
The last 12 trials in block 4 were single trials to measure the final individual tracking error of each participants in the visuomotor rotation.
Because the first trial in the visuomotor rotation was a connected trial, we could not measure the participants' \emph{individual} initial tracking error ($E_{s,0}$) on initial exposure to the visuomotor rotation as we did for the intermittent interaction and stiff interaction groups. Therefore, for the continuous interaction group only, we introduced the visuomotor rotation in one single trial in block 1 (trial 13) to measure the individual initial tracking error ($E_{s,0}$). 
We chose trial 13 in block 1 based on pilot tests. We found that the single tracking errors stabilised quickly and remained relatively constant in the trials before trial 13 (see Fig.~\ref{fig:learning-learningcurves-all}). Eight single trials with no visuomotor rotation after trial 13 (trials 14--21) were sufficient to wash out after effects of the visuomotor rotation in trial 13 before block 2 started. The continuous interaction group were coupled with a coupling stiffness of $k_s=\SI{120}{\newton\per\meter}$, same as the int. int. group.

\subsection*{Analysis}
Motor performance is analysed as the tracking error $E$, which is calculated as the root mean square of the distance between the target and their cursor of the last \SI{20}{\second} of each \SI{23}{\second} trial. We primarily focused our analysis on the single trials in the visuomotor rotation blocks, unless explicitly stated otherwise.
The tracking errors in the single and connected trials are denoted by $E_{s}$ and $E_{c}$, respectively. The tracking error in the first single trial in the visuomotor rotation blocks is referred to as the initial tracking error $E_{s,0}$.

We studied each participant's individual adaptation to the visuomotor rotation by analysing their absolute tracking error improvement and improvement rate in the single trials.
Each participant's single tracking error improvement $I_s$ was calculated as the difference between the initial tracking error $E_{s,0}$ and the final single tracking error $E_{s,f}$: $I_s = \left(E_{s,0}-E_{s,f}\right)$. 
We defined each participant's final tracking performance $E_{s,f}$ as their mean tracking performance of the last five single trials in block 4. 
Differences in individual improvement across groups were tested using a linear mixed-effect model with improvement as dependent variable, interaction group $G$ as fixed effect and pair as random variable using maximum likelihood estimation. 

Short-term visuomotor adaptation has been shown to consist of slow- and fast-adaptation processes \cite{smith_interacting_2006}. 
We fitted a function with two exponents -- one with a fast improvement rate $\lambda_f$ and another with a slow improvement rate $\lambda_s$, $\lambda_s < \lambda_f$ -- to the single trials of the visuomotor rotation blocks of each participant:
\begin{equation}
    E_{s} = a  + b_s e^{-\lambda_s (t-1)} + b_f e^{-\lambda_f (t-1)}\textrm{,}\\
	\label{eq:learning-expfunction}
\end{equation}
where $t$ is the trial number and $a$, $b_s$ and $b_f$ are constants.
The effect of interaction group on slow and fast improvement rates was tested by fitting a linear mixed-effects model with the log-transform of either the slow or fast improvement rate as dependent variable, interaction group $G$ as fixed effect and participant pair as random factor using maximum likelihood estimation. Improvement rates were log-transformed to yield improved residual distributions, which we assessed through visual inspection of the histograms and QQ-plots.

Haptic interaction with an initially more skilled partner could facilitate a participant's individual improvement and improvement rates\cite{MarchalCrespo:2013fp}. We analysed whether the difference in initial skill level between the partners impacted their individual improvement ($I_s$) and individual improvement rates ($\lambda_s$, $\lambda_f$). 
The initial tracking error $E_{s,0}$ was used as a representation of each participant's single initial skill level.
The partner's relative initial tracking error was calculated for each participant per pair as $\Delta E^p_{s,0}=\left(E_{s,0}-E^p_{s,0}\right)/E_{s,0}$, where $E_{s,0}$ is the participant's initial tracking error and $E^p_{s,0}$ is their partner's initial tracking error.
A positive $\Delta E^p_{s,0}$ means that the partner's initial tracking error was lower than the participant's own initial tracking error (i.e. the partner's initial tracking performance was better). A negative $\Delta E^p_{s,0}$ indicates that the partner's initial tracking error was worse than the participant's initial tracking error (i.e. the participant was initially better). 
To assess whether individual improvement changed significantly with interaction group $G$ and $\Delta E^p_{s,0}$, we fitted a linear mixed-effects model with individual improvement $I_s$ as dependent variable, $\Delta E^p_{s,0}$ as co-variate, interaction group $G$ as fixed effect and pair as random variable.
The improvement rates are also fitted with the same linear mixed-effects model with either $\log(\lambda_s)$ or $\log(\lambda_f)$ as dependent variable.  
All linear mixed-effects models were fitted using maximum likelihood.

Lastly, previous work showed that a participant's tracking error in connected trials $(E_c)$ depended on their partner's relative tracking error during the subsequent single trials\cite{Ganesh:2014sr,Beckers:2018bi,Takagi:2017kv}. To corroborate these results and check whether our interaction paradigm yields similar results, we calculated each participant's instantaneous improvement in tracking error \emph{during} haptic interaction as $\Delta E_c = \left(E_s-E_c\right) / E_s$, where $E_c$ is the tracking error in the connected trial and $E_s$ is the tracking error in the immediately following single trial. 
The connected trial tracking error $E_c$ could be different for the partners in a pair because of the compliant connection.
A participant's instantaneous improvement in a connected trial $\Delta E_c$ is compared to their partner's relative tracking error in the single trial that immediately followed the connected trial: $\Delta E^p_s = \left(E_s-E^{p}_s\right) / E_s$, where $E^{p}_s$ is their partner's single tracking error and $E_s$ is the participant's tracking error in the same single trial. A positive $\Delta E^p_s$ means that the partner's performance was better and a negative $\Delta E^p_{s}$ indicates that the partner's single tracking performance was worse than the participant's own single trial performance. $\Delta E_c$ and $\Delta E_s$ are only calculated for the int. int. and stiff int. group because the calculation requires pairs of connected and single trials (i.e., CS). Note that $\Delta E^p_s$, which we calculate for each pair of CS trials in the visuomotor rotation blocks, is different from $\Delta E^p_{s,0}$, which we only calculate using the first single trial in the visuomotor rotation blocks.

To test the effect of relative partner performance and group on tracking error improvement in a connected trial, we fitted a linear mixed-effect model using maximum likelihood with $\Delta E_c$ as dependent variable, interaction group $G$ (with int. int. and stiff int. group) and $\Delta E^p_s$ and the quadratic term $(\Delta E^p_s)^2$ as predictors and pair $i$ as random variable to the data (following similar studies\cite{Ganesh:2014sr,Takagi:2017kv}):
\begin{equation}
	\Delta E_c = \beta_0 +\beta_1\,\Delta E^p_s + \beta_2\,(\Delta E^p_s)^2 + \beta_3\,G +\beta_4\,\big(\Delta E^p_s\times G\big) + \beta_5\,\big( (\Delta E^p_s)^2\times G\big) + \epsilon_i\textrm{,}
\end{equation}
where $\beta_{0,\ldots,5}$ are the model coefficients and $\epsilon_i$ the unexplained variation of improvement for each pair $i$. We included the square of $\Delta E^p_s$ to include the slope increase with relative partner performance observed in the data.

We found that participant 11 showed relatively high and variable tracking errors in the last 12 single trials after the connected trials compared to the other participants in the cont. int. group (see Supplementary Fig. S1d). However, the participant's tracking performance during the connected trials was similar to the other participants in the cont. int. group. This may indicate that this participant relied too much on the haptic interaction for tracking in the visuomotor rotation or that he/she was less motivated to perform the task alone. As this is the only participant in which we observed this consistently different tracking behaviour, we decided to not include participant 11 of the cont. int. group in the analysis.

%%
%% Results
%%
\section*{Results}
We investigated whether haptic interaction between two partners facilitated their individual motor performance improvement of tracking a continuously moving target -- analysed as the tracking error -- while perturbed by a visuomotor rotation of \SI{80}{\degree}. Participants first tracked the target in a baseline block without a visuomotor rotation. We then visually rotated the on-screen cursor movement clockwise with \SI{80}{\degree} with respect to the participant's actual hand movement in the following three blocks.
Participants were na\"ive to the visuomotor rotation, which initially resulted in high tracking errors (i.e., low tracking performance) that reduced (i.e., improved tracking performance) with practice of tracking the target in the visuomotor rotation. The partners either performed the tracking task alone in single trails or haptically coupled through a dual-robot interface in connected trials.
We tested four groups to study whether haptic interaction improved individual tracking performance in the visuomotor rotation: (1) a baseline group who performed the tracking task alone (solo group), (2) a group that intermittently interacted (alternating single and connected trials; int. int. group) through a compliant connection ($k_s=\SI{120}{\newton\per\meter}$), (3) a group that also intermittently interacted through with a stiffer coupling ($k_s=\SI{250}{\newton\per\meter}$) than the int. int. group to study the effect of interaction strength on individual motor improvement (stiff int. group), and (4) a group that continuously interacted through a compliant connection ($k_s=\SI{120}{\newton\per\meter}$) in all visuomotor rotation trials to study the effect of more interaction time on individual motor improvement (cont. int. group). 
Our analysis focused on the participants' individual tracking performance improvement in the single trials in the visuomotor rotation blocks, unless explicitly stated otherwise.
\begin{figure}[h]
	\centering
	% \tikzsetnextfilename{Figure-2-Beckers}
	% \input{figures/tikz-src/learning-curves-single.tex}
	\includegraphics[]{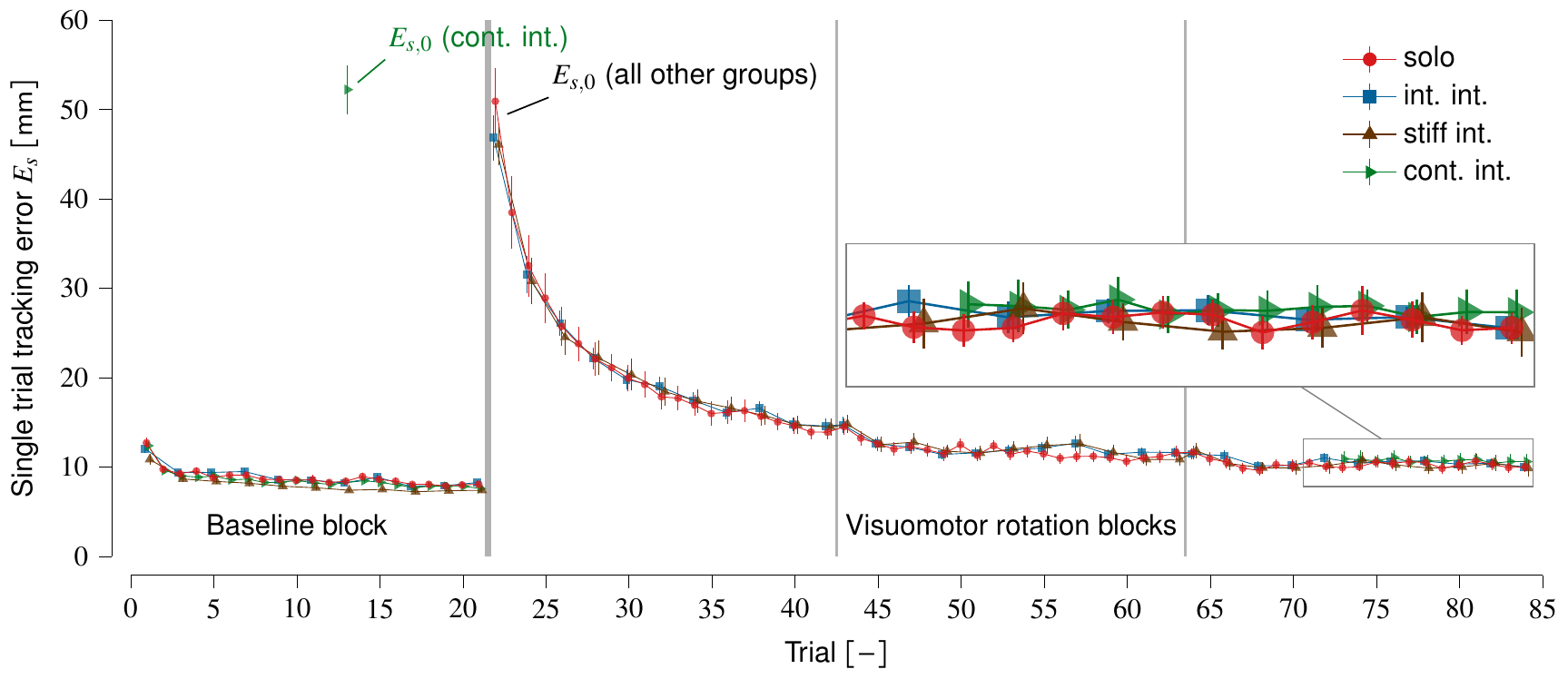}
	\caption{Tracking error in the single trials ($E_s$) in the baseline and visuomotor rotation blocks (group mean$\pm$s.e.m.). The initial tracking errors in the first single trial in the visuomotor rotation ($E_{s,0}$) for all groups are explicitly labelled; note that the cont. int. group performed one single trial with visuomotor rotation in the baseline block. The vertical lines delineate the blocks.}
	\label{fig:learning-learningcurves-all}
\end{figure}
\subsection*{Haptic human-human interaction does not yield more motor improvement or faster improvement rates}
All groups show clear and similar single trial tracking error ($E_s$) improvements in the visuomotor rotation blocks (Fig.~\ref{fig:learning-learningcurves-all}). A visual inspection of the data shows only small differences in tracking errors between groups. 
Before analysing the participant's individual improvement in the visuomotor rotation, we first checked whether the introduction of the visuomotor rotation initially increased single tracking errors similarly for all groups. 
We calculated the initial increase in tracking error using the last single trial in block 1 (without visuomotor rotation) and first single trial in block 2 (with visuomotor rotation) for the solo, int. int. and stiff int. groups. 
For the cont. int. group we calculated the increase in tracking error between trial 12 (single trial without visuomotor rotation) and trial 13 (single trial with visuomotor rotation) in block 1. 
Overall, we found no significant differences in the initial tracking error increase due to the visuomotor rotation between groups ($\chi^2(3)=2.90$, $p=0.407$).

\begin{figure}[h]
	\centering
	% \tikzsetnextfilename{Figure-3-Beckers}
	% \begin{tikzpicture}
	% 	\node[inner sep=0pt] at (0,4) {\parbox{3em}{\subcaption{}\label{fig:learning-improvement-group}}};
	% 	\node[anchor=north west] at (0,4.5) {\input{figures/tikz-src/learning-improvement-group}};
	% 	% improvement after learning
	% 	\node[inner sep=0pt] at (6.5,4) {\parbox{3em}{\subcaption{}\label{fig:learning-rates-group}}};
	% 	\node[anchor=north west] at (6.5,4.5) {\input{figures/tikz-src/learning-rates}};
	% \end{tikzpicture}
	\includegraphics{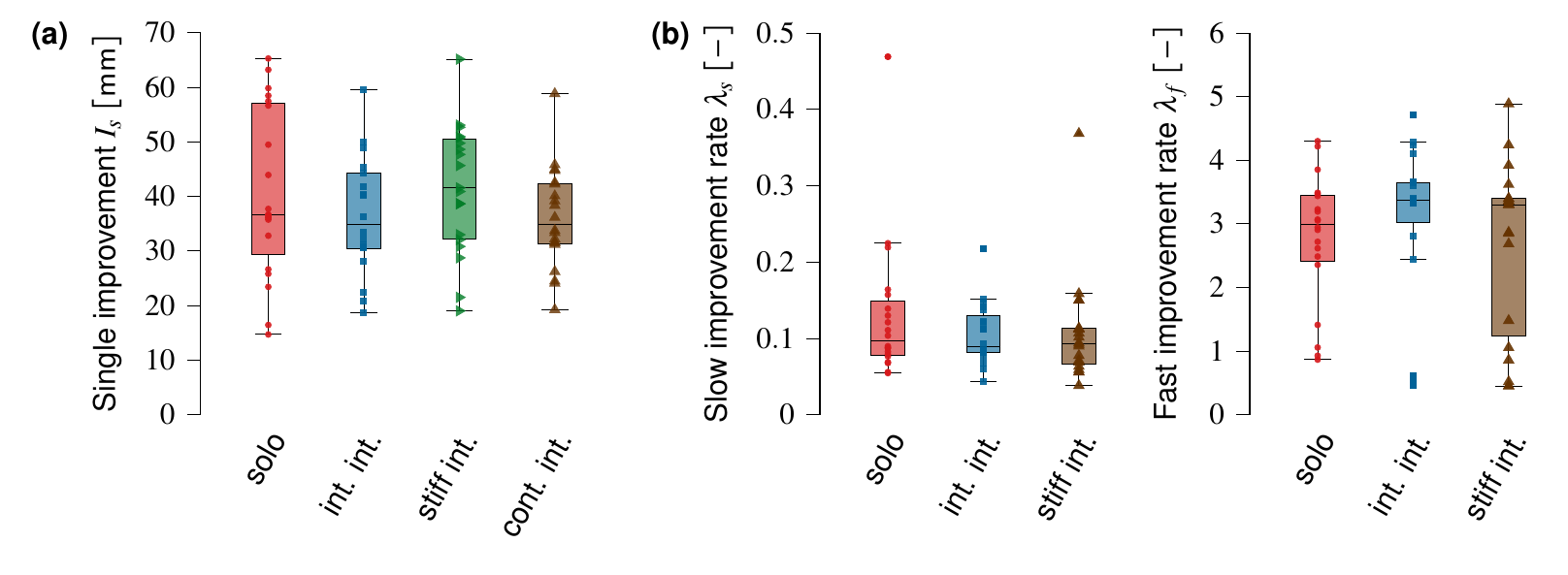}
	\caption{Tracking error improvement and improvement rates in the visuomotor rotation blocks.
	\textbf{(a)} Single tracking error improvement for all groups.
	\textbf{(b)} Individual slow $\left(\lambda_s\right)$ and fast $\left(\lambda_f\right)$ improvement rates for the solo, intermittent interaction, and stiff interaction groups. While we used the log-transform of learning rates, we plotted the non-transformed data here. The markers show the participants' individual data.}
	\label{fig:learning-group-improvement-learning}
\end{figure}

To analyse the tracking error improvement curves in the visuomotor rotation blocks, we calculated each participant's \emph{individual} single tracking error improvement $I_s$ and each participant's slow and fast improvement rates based on their single trials ($\lambda_s$ and $\lambda_f$, respectively). We found no significant differences in individual tracking error improvement between groups (Fig.~\ref{fig:learning-group-improvement-learning}a, effect of group G on improvement: $\chi^2(3)=3.22$, $p=0.360$). These results indicate that haptic interaction does not result in more individual performance improvement compared to practising the task solo.

In addition, haptic interaction did not result in significantly different slow and fast improvement rates for the solo, intermittent interaction and stiff interaction groups, see Fig.~\ref{fig:learning-group-improvement-learning}b (effect of group $G$ on slow improvement rate: $\chi^2(2)=1.09$, $p=0.582$ and fast improvement rate: $\chi^2(2)=2.30$, $p=0.316$). 
Because the participants in the continuous interaction group did not perform any single trials in the early stages while adapting to the visuomotor rotation, we could not extract their individual motor improvement rates. 

\subsection*{The effect of the partner's relative initial skill level on individual improvement and improvement rates}
Although we found no significant effect of haptic interaction on individual improvement on a group level, some participants could have benefited from the haptic interaction depending on the difference in initial skill level between the partners in a pair (i.e., the difference between $E_{s,0}$ of both partners) \cite{Guadagnoli:2004ig, MarchalCrespo:2015fa, AvilaMireles:2017cf}. 
For example, haptic interaction with an initially highly skilled partner could result in a higher individual improvement $I_s$ compared to interaction with a similarly skilled partner or compared to a participant who practised the task without interaction with a better partner. 

\begin{figure}[!ht]
	\centering
	% \tikzsetnextfilename{Figure-4-Beckers}
	% \begin{tikzpicture}
	% 	\node[inner sep=0pt] at (0,4) {\parbox{3em}{\subcaption{}\label{fig:learning-Es0rel-improvement-rates-a}}};
	% 	\node[anchor=north west] at (-0.3,4.3) {\input{figures/tikz-src/relperf-learning-rates-a}};
	% 	% improvement after learning
	% 	\node[inner sep=0pt] at (5.5,4) {\parbox{3em}{\subcaption{}\label{fig:learning-Es0rel-improvement-rates-b}}};
	% 	\node[anchor=north west] at (5.2,4.3) {\input{figures/tikz-src/relperf-learning-rates-b}};
	% \end{tikzpicture}
	\includegraphics{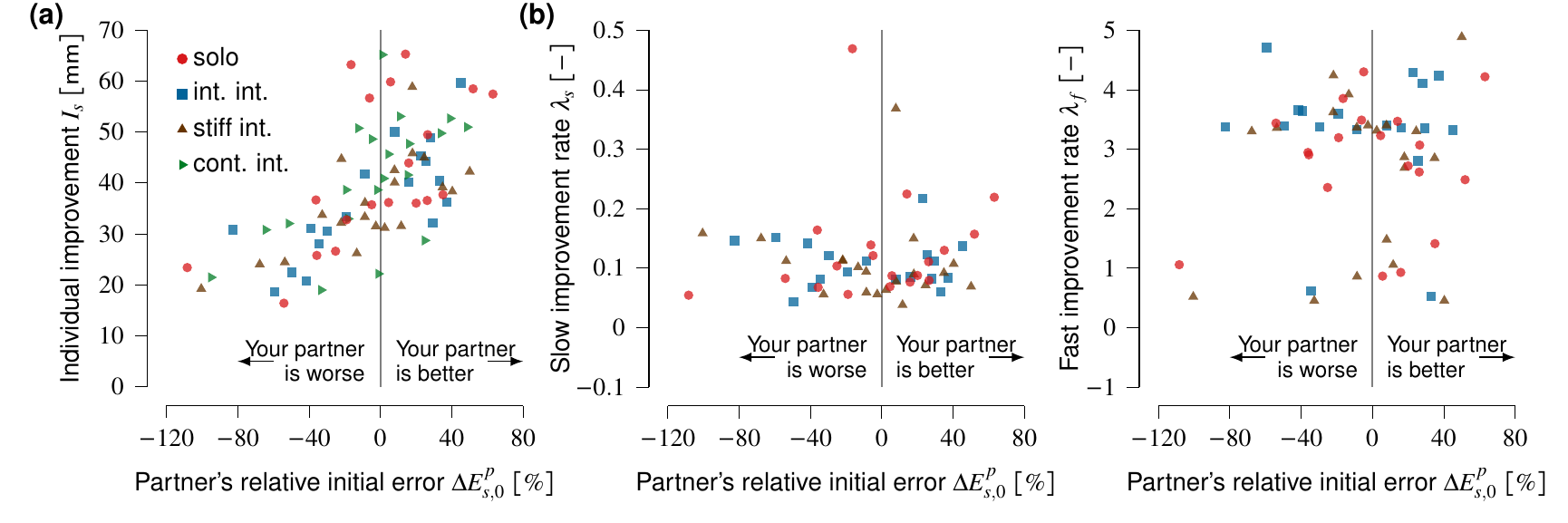}
	\caption{The partner's relative initial tracking error $\Delta E^p_{s,0}=\left(E_{s,0}-E^p_{s,0}\right) / E_{s,0}$ versus \textbf{(a)} individual improvement $I_s$ and \textbf{(b)} individual improvement rates ($\lambda_s$ and $\lambda_f$). We refer to the two solo participants in each pair as partners for consistency with the pairs in the interaction groups, although they were never connected.}
	\label{fig:learning-Es0rel-improvement-rates}
\end{figure}

To test whether haptic interaction benefited individual improvement on top of the potential effect of the initial skill difference between partners, we need to compare the interaction groups to participants who never interacted. 
The analysis focuses on the partner's relative initial error in the first single trial $(\Delta E^p_{s,0})$ before any connected trials occurred in the visuomotor rotation blocks, and compares $\Delta E^p_{s,0}$ with the overall individual improvement ($I_s$, $\lambda_s$, and $\lambda_f$). 
We can, therefore, include the solo group as a control group that never had connected trials by calculating $\Delta E^p_{s,0}$ for the solo pairs and comparing it to each solo participant's individual improvement and improvement rates.  
Note that although the solo participants performed the experiment in pairs, they are not considered to be partners, as they were never connected and were thus not able to influence each other (in principle we could even randomly select pairs of solo participants for the analysis). 
For consistency with the interaction groups, we still refer to the solo participants in a pair as partners. 
If haptic interaction has an additional effect on individual improvement on top of the partner's relative initial skill level, we would expect a difference between the solo group and the interaction groups, depending on the skill difference.
Fig.~\ref{fig:learning-Es0rel-improvement-rates} shows each participant's individual improvement $I_s$ and improvement rates ($\lambda_s$ and $\lambda_f$) versus their partner's relative initial error $\Delta E^p_{s,0}$.

A participant's individual improvement $I_s$ depended significantly on their partner's relative initial error $\Delta E^p_{s,0}$ ($\chi^2(1)=121.07$, $p<10^{-6}$, see Fig.~\ref{fig:learning-Es0rel-improvement-rates}a). This indicates that the more skilled their partner initially was compared to their own initial skill level, the higher the participant's own individual improvement $I_s$. We found no significant differences between groups $G$ ($\chi^2(3)=3.27$, $p=0.352$). The interaction effect ($\Delta E^p_{s,0}\times G$) was also not significant ($\chi^2(3)=6.69$, $p=0.083$). Hence, individual improvement depends significantly on the partner's relative initial performance, and this effect is similar for all groups including the solo group.

The improvement rates did not depend on the partner's relative initial error ($\Delta E^p_{s,0}$) or interaction group $G$ (Fig.~\ref{fig:learning-Es0rel-improvement-rates}b).
The slow improvement rate ($\lambda_s$) neither depended on the partner's relative initial performance  ($\chi^2(1)=0.33$, $p=0.564$) nor on the interaction group ($\chi^2(2)=1.12$, $p=0.570$). The interaction effect $(\Delta E^p_{s,0} \times G)$ was also not significant ($\chi^2(2)=5.08$, $p=0.078$). 
Similarly, we found no effect of partner's relative initial performance on the fast improvement rate ($\lambda_f$; $\chi^2(1)=0.22$, $p=0.639$) and the fast improvement rate did not depend on interaction group ($\chi^2(2)=2.57$, $p=0.277$). The interaction effect $(\Delta E^p_{s,0} \times G)$ was also not significant ($\chi^2(2)=1.07$, $p=0.586$) for the fast improvement rate.

\subsection*{Tracking performance improves during haptic interaction depending on the partner's relative single tracking performance}
A common finding in similar haptic human-human interaction studies is that haptic interaction improves individual tracking error \emph{during} interaction ($\Delta E_c$) depending on the partner's relative tracking error in the subsequent single trial ($\Delta E^p_s$) \cite{Ganesh:2014sr,Reed:2008ie,Beckers:2018bi,Takagi:2018ku}. Although our focus in this paper is on analysing single trial improvement, we also analysed $\Delta E_c$ versus $\Delta E^p_s$ for the int. int. and stiff int. groups to check whether our data corroborates previous studies.
Note that this section is different from the analysis in the previous section, in which we analysed \emph{individual} improvement $I_s$ based on \emph{single trials} only for all groups, whereas here, we analyse the immediate effect of haptic interaction during the \emph{connected trials} for the int. int. and stiff int. groups only. Another difference is that here, $\Delta E^p_s$ is calculated for each single trial after a connected trial in the visuomotor rotation blocks, which is different from the partner's relative \emph{initial} error $\Delta E^p_{s,0}$ used in the previous section, which is only calculated using the first single trial in the visuomotor rotation blocks.

\begin{figure}[!ht]
	\centering
	% \tikzsetnextfilename{Figure-5-Beckers}
	% \input{figures/tikz-src/learning-hhi-improvement-all}
	\includegraphics[]{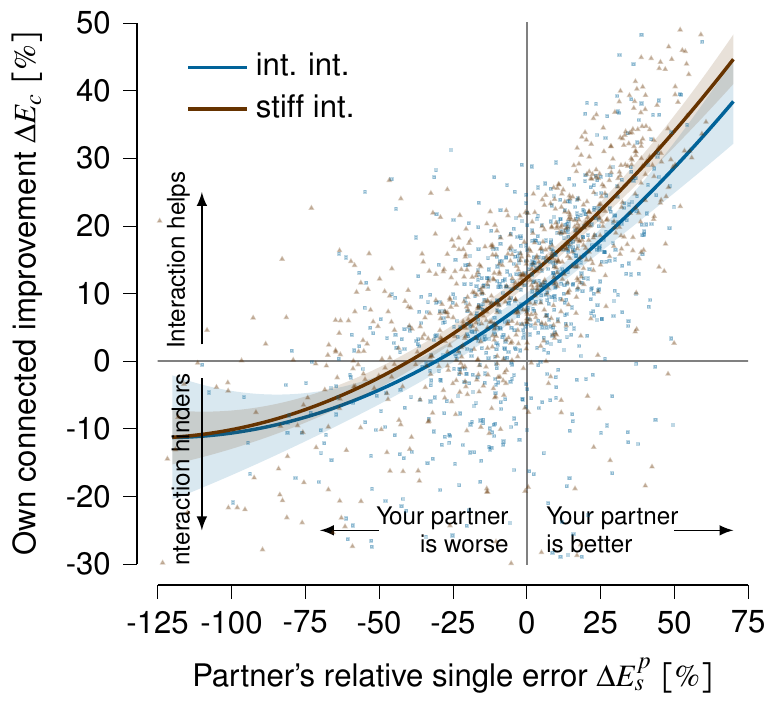}
	\caption{Tracking error improvement during interaction in a connected trial ($\Delta E_c = \left(E_s-E_c\right) / E_s$) versus relative partner performance in the subsequent single trial ($\Delta E^p_s = \left(E_s-E^p_{s}\right)/E_s$) for the int. int. and stiff int. groups.}
	\label{fig:interaction-improv-relperf}
\end{figure}

The partner's relative error and interaction group (int. int. or stiff int. group) had a significant effect on the tracking error improvement during interaction $\Delta E_c$ (see Fig.~\ref{fig:interaction-improv-relperf}). 
Specifically, the quadratic term $(\Delta E^p_s)^2$ (see Eq.~\ref{fig:interaction-improv-relperf}) had a significant effect on connected tracking improvement ($\chi^2(1)=5.82$, $p=0.016$), indicating that the connected improvement increased with a greater slope rise with a progressively better partner. 
The interaction effect $\big( (\Delta E^p_s)^2\times G\big)$ was also significant ($\chi^2(1)=3.98$, $p=0.046$), indicating that $(\Delta E^p_s)^2$ had a significantly larger effect on $\Delta E_c$ for the stiff int. group compared to the int. int. group. 
In other words, interacting through a stiffer coupling resulted in progressively more connected improvement compared to a weaker coupling, in particular for participants whose partners were better at the task (participants with $\Delta E^p_s > \SI{0}{\percent}$).

Our results also showed that interaction can hinder tracking performance compared to single performance (i.e. $\Delta E_c < \SI{0}{\percent}$). When interacting with a partner whose relative single error was approximately $\Delta E^p_s < \SI{-30}{\percent}$ for the int. int. group and $\Delta E^p_s < \SI{-39}{\percent}$ for the stiff int. group, the fitted model predicted that interaction would hinder performance. 
Interestingly, even interaction with a partner who was slightly worse at the task (e.g. $\SI{-30}{\percent} \leq \Delta E^p_s \leq \SI{0}{\percent}$ for the int. int. group and $\SI{-39}{\percent} \leq \Delta E^p_s \leq \SI{0}{\percent}$ for stiff int.) still resulted in an improvement during interaction ($\Delta E_c \geq \SI{0}{\percent}$).

%%
%% Discussion
%%
\section*{Discussion}

We investigated whether haptic interaction between two humans performing the same target tracking task in a visuomotor rotation perturbation enhanced their individual motor adaptation in terms of motor improvement and improvement rate compared to someone who practised the task alone.
This work was motivated by the results of Ganesh et al.\cite{Ganesh:2014sr}, who found a significant benefit of haptic interaction on individual motor improvement in the same motor task. Here, we repeated their study using the same interaction paradigm and motor adaptation task. In addition to repeating Ganesh et al.\cite{Ganesh:2014sr}, in which we compared a group in which partners intermittently interacted with a solo group, we also added a group who spent more time interacting and a group who intermittently interacted through a stronger haptic connection.
In contrast to Ganesh et al.\cite{Ganesh:2014sr}, we found no effect of intermittent interaction on individual improvement or improvement rate compared to the solo group. More interaction time (e.g. more connected trials) to allow the participants to benefit more from the interaction -- if those benefits would have been present in our study -- did not improve individual improvement either. Interaction through a stronger connection also resulted in similar individual improvement. Although we found an effect of the partner's initial performance level on individual improvement, there was no difference between the interaction and solo groups. Improvement rates did not depend on the partner's relative skill level.

Our results are in line with observations in motor adaptation literature and robot-assisted motor skill acquisition studies. A consistent finding in this field is that movement errors drive motor adaptation; reducing movement errors, for example through haptic assistance, does generally not facilitate individual motor improvement \cite{MarchalCrespo:2013fp,Shadmehr:2010fi,Wei:2009bi}.
In accordance with other haptic human-human interaction studies \cite{Ganesh:2014sr,Takagi:2017kv,Takagi:2018ku,AvilaMireles:2017cf,kager2019effect,Beckers:2018bi}, we found that tracking errors generally were smaller \emph{during} haptic interaction ($\Delta E_c > 0$) depending on the partner's relative single trial error ($\Delta E^p_s$), specifically for participants whose partners had better single tracking performance.  
Haptic interaction could be seen as a compliant guidance that allows each participant to independently perform the task, but still benefit from the error-correcting guidance of their partner. This gives the interacting partners an incorrect good impression of their tracking performance, but reduces tracking error as driving training signal for their own motor adaptation \cite{MarchalCrespo:2013fp}. This observation is supported by several studies that showed that robot-generated haptic guidance temporarily improved motor performance while the participant received the guidance, but did not improve subsequent individual motor skill acquisition \cite{Winstein:1994gf,vanAsseldonk:2009do,OMalley:2006dv,MarchalCrespo:2014ec,Heuer:2015tb}.

Although the group-based analysis found no benefit of haptic interaction on individual improvement, we also investigated whether some participants could have benefited from the interaction depending on their partner's relative initial skill level.
Avila-Mireles et al.\cite{AvilaMireles:2017cf} found that haptic interaction between two partners who were na\"ive to the task resulted in more individual motor skill acquisition benefits compared to interaction with an expert. Interaction with an expert only helped if the novice had prior solo experience in the task.
Kager et al.\cite{kager2019effect} found no effect of partner skill level on individual tracking skill improvement in a tracking task similar to ours. 
We did not control the initial skill difference between the partners, resulting in a range of relative skill differences.
We found that haptic interaction with an initially more skilled partner resulted in higher individual improvement. 
However, this effect could also be because participants with a high initial single tracking error -- who are more likely to interact with a superior-performing partner -- had more room to improve. This might be related to the task and further research is warranted. We used the skill difference and not each participant's own absolute initial tracking error, as the skill difference of the partner could have impacted to what extent haptic interaction helped or even hindered individual improvement. 
Lastly, and more importantly, we found no additional benefit of haptic interaction for individual improvement on top of the partner's initial skill level compared to solo participants (the effect of group $G$ and interaction effect $(\Delta E^p_{s,0} \times G)$ on $I_s$ were not significant). Hence, haptic interaction with a more-skilled or less-skilled partner does not benefit individual improvement compared to practising the task alone, supporting our group-based conclusion.

%% DIFFERENCES (AND SIMILARITIES) GANESH STUDY
Although we used the same interaction paradigm and motor task as Ganesh et al.\cite{Ganesh:2014sr}, a number of differences in the experiment design and analysis could explain the different results. First, our solo and intermittent interaction groups used double the number of participants compared to the solo and interaction groups of Ganesh et al.\cite{Ganesh:2014sr} (twenty participants or ten pairs per group in our study, compared to ten participants or five pairs). In addition, we tested another forty participants -- albeit with slight variations of the interaction paradigm -- without any effect on individual improvement. 

Second, we calculated each participant's own single improvement $I_s$ on their own initial tracking error $E_{s,0}$ and own final tracking error $E_{s,f}$, whereas Ganesh et al.\cite{Ganesh:2014sr} based improvement on the group-mean of the initial tracking error (see Fig 1c in their paper). As our data showed a reasonable spread in initial tracking error between participants indicating differences in initial skill levels (see Supplementary Fig. S1), we believe that calculating individual improvement based on each individual's initial and final tracking performance is a more precise representation for each individual's improvement. 
In fact, if we would have calculated and analysed each participant's improvement based on their group-average initial tracking error, we would find that the solo group improved \emph{more} than the intermittent and stiff interaction groups and similar to the continuous interaction group. Hence, our conclusion would have been that intermittent haptic interaction would even \emph{impede} motor improvement compared to practising the task solo (see Supplementary Fig. S2).

Third, we used an alternating sequence of single and connected trials for the intermittent interaction group, whereas Ganesh et al.\cite{Ganesh:2014sr} used a semi-randomised sequence. However, we believe that our trial sequence had little effect on the participant's individual tracking improvement, based on a direct comparison of the solo and continuous interaction groups. These two groups had two opposite extremes of the possible trial sequences; either all single or all connected during the initial adaptation phases (blocks 2 and 3). The solo and continuous interaction groups showed no significant difference in improvement compared to the other groups, indicating that the trial sequence or amount of interaction time in our task did not seem to significantly affect individual improvement.

Lastly, our robot interface's workspace (diameter of \SI{20}{\centi\meter}) was smaller than that of Ganesh et al.\cite{Ganesh:2014sr} (diameter of \SI{30}{\centi\meter}) due to hardware limitations. As a result, we had to design a different target trajectory that required a smaller range of motion. This resulted in smaller improvement in tracking errors in the single and connected trials compared to Ganesh et al.\cite{Ganesh:2014sr}. Still, our data showed prominent yet similar individual improvement curves for all groups (see Fig.~\ref{fig:learning-learningcurves-all}) and we found significant improvement of tracking error during interaction in the connected trials similar to Ganesh et al.\cite{Ganesh:2014sr} and other studies using the same interaction paradigm \cite{Takagi:2017kv,Beckers:2018bi,Takagi:2018ku} (Fig.~\ref{fig:interaction-improv-relperf}).

Although the focus of this paper is on the effect of haptic interaction on individual motor improvement, we also analysed the tracking error improvement during interaction in the connected trials compared to the subsequent single trials (denoted by $\Delta E_c$) to check whether the haptic interaction paradigm yielded similar result as previous haptic human-human interaction studies work (specifically \cite{Takagi:2018ku, Beckers:2018bi, Ganesh:2014sr}). The extent to which haptic interaction helps ($\Delta E_c > \SI{0}{\percent}$) or hinders ($\Delta E_c < \SI{0}{\percent}$) tracking performance depends on the partner's relative single performance $E^p_s$. Interaction with a better partner improved tracking performance. Interestingly, even when a partner's relative single error was slightly worse (e.g., $\SI{-30}{\percent}<\Delta E^p_s<\SI{0}{\percent}$), haptic interaction still improved performance during interaction. Interaction hindered tracking performance for a partner relative error of  $\Delta E^p_s < \SI{-30}{\percent}$. These observations corroborate previous studies \cite{Takagi:2018ku, Beckers:2018bi, Ganesh:2014sr}. 
Lastly, our observation that a stiffer interaction yield progressively more tracking error benefit is also observed in previous work \cite{Takagi:2018ku}.

It is likely that the benefits of haptic human-human interaction on motor skill acquisition are task- and instruction-dependent. For example, Avila-Mireles et al.\cite{AvilaMireles:2017cf} found that a uni-manual skill learned during haptic interaction with a human partner could benefit the participant's skill level in the same task that the participant would now perform bi-manually. 
We only investigated short-term motor improvement over a \SI{2}{\hour} session; the effects of haptic human-human interaction on long-term skill acquisition and retention remain unknown. Lastly, we neither made the participants explicitly aware of the connection, nor assigned roles, such as teacher-student. Assigning roles or making partners aware of the connection could influence their interaction strategies and motor improvement of the partners\cite{Granados:2007wh}.

In conclusion, our study found no benefit of haptic interaction between partners on short-term individual motor improvement in a tracking task with a large visuomotor rotation. We could not corroborate the findings of Ganesh et al.\cite{Ganesh:2014sr} on individual motor improvement.
Recalling the example of a physiotherapist assisting a patient to regain motor abilities, we are aware of the limited generalisability of our in-the-lab-results to realistic motor tasks. Haptic human-human interaction likely plays an important role in motor skill acquisition in our lives. Even though we studied three aspects of haptic human-human interaction in our specific tracking task -- interaction time, interaction strength and partner skill level -- further investigation into the mechanisms behind the possible benefits of haptic human-human interaction is warranted. Furthermore, as we minimised any social interaction between partner except through the interaction force, we need to take important aspects of joint action such as (conscious) motor coordination and role distribution into account in future studies.

\section*{Data availability}
Data and analysis scripts that support the findings of this study are available through the following link: \url{https://doi.org/10.4121/uuid:a4f1add7-892d-456e-8b8a-75a708d01f15}.

\section*{Acknowledgements}
We thank Jeanine van Bruggen, Frits Wiersma and Ingrid van den Heuvel for their assistance with the experiments. We also thank Jared Atkinson and Victor Sluiter for their insightful comments on the manuscript. This research is supported by the Netherlands Organisation for Scientific Research (NWO), project no. 13524.

\section*{Author contributions statement}
N.B. conceived the experiments, conducted the experiments and wrote the manuscript. All authors analysed and discussed the results and reviewed the manuscript. 

\section*{Additional information}

\textbf{Competing interests} The authors declare no competing interests. 

% \bibliography{bibliography}

\end{document}